# Two-Dimensional TaSe$_2$ Metallic Crystals:
# Spin-Orbit Scattering Length and Breakdown Current Density

Adam T. Neal, Yuchen Du, Han Liu, Peide D. Ye*

*School of Electrical and Computer Engineering and Birck Nanotechnology Center,*

*Purdue University, West Lafayette, IN 47907, USA*

*correspondence to: yep@purdue.edu

**Abstract**

We have determined the spin-orbit scattering length of two-dimensional layered 2H-TaSe$_2$ metallic crystals by detailed characterization of the weak anti-localization phenomena in this strong spin-orbit interaction material. By fitting the observed magneto-conductivity, the spin-orbit scattering length for 2H-TaSe$_2$ is determined to be 17 nm in the few-layer films. This small spin-orbit scattering length is comparable to that of Pt, which is widely used to study the spin Hall effect, and indicates the potential of TaSe$_2$ for use in spin Hall effect devices. In addition to strong spin-orbit coupling, a material must also support large charge currents to achieve spin-transfer-torque *via* the spin Hall effect. Therefore, we have characterized the room temperature breakdown current density of TaSe$_2$ in air, where the best breakdown current density reaches $3.7\times10^7$ A/cm$^2$. This large breakdown current further indicates the potential of TaSe$_2$ for use in spin-torque devices and two-dimensional device interconnect applications.

Although studied for some time,[1,2] the transition metal dichalcogenide (TMD) family of materials has attracted increased attention in the nanoelectronics community due to their two-dimensional layered structure, following the prolific research into graphene.[3-6] With graphene's zero bandgap limitation for transistor technology, much of the nanoelectronics community's interest in TMDs has been focused on the semiconductors, particularly $MoS_2$, with the demonstration of single-layer and few-layer field-effect transistors.[7-9] Metallic TMDs, on the other hand, have received much less attention in the nanoelectronics community thus far, but recent works on exfoliated $TaSe_2$ indicate that interest is on the rise.[10-12] Notably, single-layer $TaSe_2$ has been recently characterized by Raman spectroscopy.[13] Historically, metallic TMDs have been intensely studied by material physicists and condensed matter physicists due to their superconducting[1,2] and charge density wave[14] properties, which remain active areas of research.

Obviously, metallic TMDs are not suitable for field-effect transistor channel materials, similar to graphene. One possible nanoelectronics application, previously proposed for graphene,[15] is the use of metallic TMDs as device interconnects for an all two-dimensional (2D) material logic technology. Another application of metallic TMDs, particularly of 2H-$TaSe_2$ on which we will focus in this work, is in spintronics devices. Angle-resolved photoemission spectroscopy (ARPES) measurements of 2H-$TaSe_2$ reveal a "dog-bone" like structure of the Fermi surface in the "normal" (not charge density wave) state,[16,17] and this structure is attributed to strong spin-orbit coupling in $TaSe_2$.[18] This strong spin-orbit coupling may make $TaSe_2$ an ideal 2D material for generation of spin currents *via* the spin Hall effect. Motivated by these potential applications of 2H-$TaSe_2$, we determine, for the first time, the spin-orbit scattering length of $TaSe_2$ by characterizing the weak anti-localization phenomena in the material. In addition to strong spin-orbit coupling, a material must also support large charge currents to achieve spin-transfer-torque *via* the spin Hall effect. The ability to conduct large charge currents is also important for the previously mentioned two-dimensional interconnect application. Therefore, we have also characterized the breakdown current density of 2D TaSe$_2$ crystals for the first time.

**Results and Discussion**

First of all, it is important to establish the polytype of the TaSe$_2$ samples used in this work.  The 1T polytype, with octahedral coordination, and the 2H polytype, with trigonal prismatic coordination, are the two most studied in the literature.  For the 1T polytype of TaSe$_2$, the material is in the incommensurate charge density wave state below 600 K and in the commensurate charge density wave state below 473 K.[14,19,20] The transition at 473K is accompanied by a stark discontinuity in the resistivity as a function of temperature.  The 2H polytype does not transition into the incommensurate charge density wave state until ~120K, and the commensurate charge density wave sets in below 90K.  In contrast to the 1T polytype, there are no discontinuities in the resistivity as a function of temperature, but there is a characteristic change in the slope of the resistivity *versus* temperature curve at onset of the incommensurate charge density wave state at ~120 K.[14,19,20]  These properties of the resistivity *versus* temperature allow one to distinguish between the two polytypes of TaSe$_2$ electrically.  Figure 1(d) shows the resistivity of the TaSe$_2$ used in this work as a function of temperature from 4 K to 300 K.  The characteristic change in the slope of the resistivity *versus* temperature curve, indicated by the arrow in the figure, confirms that the TaSe$_2$ used in this work is of the 2H polytype.  The crystal structure of 2H-TaSe$_2$ is shown in Figure 1 (a)-(c) with layered 2D structures as expected.

With the polytype of the TaSe$_2$ established, we determine the electrically active thickness of the flake used to study spin-orbit coupling of TaSe$_2$. Figure 2(a) shows atomic force microscopy (AFM) image of the TaSe$_2$ device with the AFM height measurement of the flake overlaid. The 2D crystal has a physical thickness of ~12 nm. A calculation of the Hall coefficient from Hall effect measurements using the thickness measured by AFM yields Hall coefficients which are much too large compared to those published in the literature.[21-23] Because of this discrepancy, we conclude that the physical thickness of the flake as measured by AFM is not electrically active, perhaps due to oxidation of the top and bottom layers of the TaSe$_2$ flake while exposed to air for long periods of time. We also note that the devices with much thinner flakes cannot be measured reliably. Considering these observations, we can estimate the electrically active thickness of our TaSe$_2$ flake as the Hall coefficients from the literature divided by the Hall slope measured for our TaSe$_2$ device, $t_{ele} = R_h / \frac{dR_{xy}}{dB}$. The Hall slope $\frac{dR_{xy}}{dB}$ is the slope of the Hall resistance, $R_{xy} = V_h/I$, as a function of magnetic field. The measured Hall slopes for this device as a function of temperature are shown in Figure 2(b). We perform this thickness estimation at two temperatures, T~5K and T~ 120K, where the Hall slope for our flake at 120K was estimated by linear extrapolation using the measured Hall slopes in Figure 2(b). The electrically active thickness, $t_{ele}$, is determined to be 0.81 nm and 0.88 nm for 5K and 120K, respectively. Therefore, from this estimation *via* the Hall effect measurement, we conclude that only one or two atomic layers of the TaSe$_2$ flake are electrically active. The dependence of the magneto-conductivity on the angle of the magnetic field, shown in Figure 2(c), also confirms this claim that the system studied is a two-dimensional electron system. The resistivity and Hall coefficient plotted in Figure 1(d) and Figure 2(b) were calculated using the electrically active thickness, $t_{ele}$, that we have determined. Note that the change sign change of the Hall coefficient is expected and is related to the reconstruction of the Fermi surface as the charge density wave state develops with decreasing temperature.[21]

We now study the spin-orbit coupling strength of 2H-TaSe$_2$. Indeed, the strong spin-orbit coupling indicated by the "dog's bone" Fermi surface shape[16,17] is confirmed by magneto-transport measurements. Figure 3(a) shows the differential sheet conductance of TaSe$_2$ as a function of magnetic field for various temperatures. TaSe$_2$ exhibits a negative magneto-conductivity, characteristic of weak anti-localization, which indicates the strong spin-orbit coupling of TaSe$_2$. A classical background has been subtracted from the data, determined by fitting the data at $T = 8$ K and $B$ more than one Tesla where the localization phenomena is suppressed. To quantitatively determine the spin-orbit scattering length, we must first determine the dimensionality of the weak anti-localization phenomena in the system. Figure 2(c) shows the differential magneto-conductivity for different angles between the magnetic field and the sample. In this case no classical background is subtracted. The angular dependence shows that the weak anti-localization phenomenon behaves two dimensionally. The differential magneto-conductivity, $\Delta\sigma$, can be described for 2D weak localization by the following equation:[24-26]

$$\Delta\sigma = n_v n_s \frac{e^2}{4\pi^2 \hbar} \left( F\left(\frac{B}{B_\phi + B_{so}}\right) + \frac{-1}{n_s}\left(F\left(\frac{B}{B_\phi}\right) - F\left(\frac{B}{B_\phi + 2B_{so}}\right)\right)\right) \quad (1)$$

$$F(z) = \psi\left(\frac{1}{2} + \frac{1}{z}\right) + \ln(z), \quad B_* = \frac{\hbar}{4eL_*^2}, \quad * = \phi, so$$

where $n_v$ is the valley degeneracy, $n_s = 2$ for the spin degeneracy, $e$ the charge of an electron, $\hbar$ is Planck's constant divided by $2\pi$, $L_\phi$ the phase coherence length, and $L_{so}$ the spin-orbit scattering length. $L_\phi$ and $L_{so}$ are the free parameters which allow fitting of the data. The number of valleys, $n_v$, can be determined from ARPES performed on TaSe$_2$ in its commensurate charge density wave state.[17,21,27] In the commensurate charge density wave state, the TaSe$_2$ lattice is deformed, effectively increasing the period of the material system in real space. This leads to a smaller Brillouin zone in k-space compared to the undeformed material and also causes reconstruction of the Fermi surface. ARPES indicates that there are three independent valley's in the charge density wave Brillouin zone, therefore we choose $n_v = 3$ when fitting the weak anti-localization data. The solid orange line in Figure 3(a) shows an example fit of the weak-anti-localization peak using Equation 1.

We can now determine the spin-orbit scattering length of TaSe$_2$ by fitting the weak anti-localization data in Figure 3(a), along with others not shown. Figure 3(b) shows the phase coherence length $L_\phi$ and the spin-orbit scattering length $L_{so}$ determined from the fittings as a function of temperature. We find that $L_{so}$ is independent of temperature, while $L_\phi$ decreases as $T^{-1}$, which is consistent with dephasing due to electron-electron scattering without too much disorder.[28] Because $L_{so}$ in Figure 3(b) is independent of temperature, we take their average and determine that the spin-orbit scattering length $L_{so} = 17$ nm for 2H-TaSe$_2$. This length is comparable to the spin-orbit scattering length of Pt ($L_{so} = 12$ nm)[29], widely used to study the spin Hall effect, and indicates the potential of TaSe$_2$ for use in 2D spintronics devices.

The weak anti-localization is also be suppressed by increasing bias current as shown in Figure 4(a). The higher current bias adds energy to the sample, increasing the electron temperature above that of the helium bath, leading to suppression of the weak anti-localization. Figure 4(b) shows the phase coherence length and spin-orbit scattering length determined from the data in Figure 4(a). The phase coherence length decreases as $I^{-1/2}$, which indicates that the electron temperature increases as $I^{+1/2}$ considering $L_\phi \propto T^{-1}$ as previously determined. The same relationship between bias current and electron temperature has also been observed for a two-dimensional electron gas in the quantum Hall regiem.[30] This further confirms that the studied TaSe$_2$ device is, electrically, an atomically-thin material system.

Finally, we evaluate the breakdown current density of 2H-TaSe$_2$ in order to determine its potential to achieve spin Hall effect based spin-transfer-torque by DC characterization of the total 18 fabricated devices. The average room temperature resistivity of the TaSe$_2$ flakes used for the breakdown current measurements was $1.9 \times 10^{-4}$ $\Omega \cdot$cm, determined using four terminal measurements of 8 of the devices. The contact resistance of the Ni/Au contact to the TaSe$_2$ flakes was also estimated by subtracting the four terminal resistance from the two terminal resistance and dividing by two. The average contact resistance determined from the 8 four terminal devices was 0.74 $\Omega \cdot$mm, which is one order of magnitude smaller compared to metal/MoS$_2$ contacts.[31] This is because TaSe$_2$ is metallic and forms Ohmic contacts with metals while MoS$_2$ is a semiconductor and forms Schottky contacts with metals. To achieve this low contact resistance, the Ni/Au contact was deposited as soon as possible after TaSe$_2$ exfoliation to minimize surface oxidation and avoid oxide barriers between the Ni/Au contact and the TaSe$_2$. Flakes were also stored in a nitrogen box before metal contact deposition to help minimize the surface oxidation. Measurement of the breakdown current density is performed by continuously increasing the bias voltage across the device until a decrease in current more than one order of magnitude is observed. Figure 5(a) shows the current density *versus* bias voltage data for the device showing the highest breakdown current density observed among the 18 devices. The breakdown current density is taken as the current density immediately before the sharp decrease in current was observed. Figure 5(b) shows the histogram of the breakdown currents determined from the 18 devices. The maximum breakdown current density observed is $3.7 \times 10^7$ A/cm$^2$, the average $1.9\times10^7$ A/cm$^2$, the minimum $0.5\times10^7$ A/cm$^2$ and the standard deviation $0.8\times10^7$ A/cm$^2$. The flake thickness as characterized by AFM was used when computing the breakdown current densities, so these reported current densities could be slightly underestimated if we consider the surface oxidation. These breakdown currents are comparable to the charge currents used to induce spin-transfer-torque *via* the spin Hall effect in Tantalum thin films,[32] indicating the possibility of TaSe$_2$ based 2D spin-torque devices. The breakdown currents are about one order of magnitude less than those of graphene;[15,33] however, they are comparable to those of MoS$_2$.[34] These large breakdown currents also indicate the potential of TaSe$_2$ as a 2D interconnect material, particularly if used in conjunction semiconducting TMDs where the similar crystal structure may provide some integration advantages.[35]

**Conclusions**

In conclusion, we have determined the spin-orbit scattering length of 2H-TaSe$_2$ by detailed characterization of the weak anti-localization phenomena in the material. By fitting the observed magneto-conductivity, the spin orbit scattering length for 2H-TaSe$_2$ is determined to be 17 nm. This small spin orbit scattering length is comparable to that of Pt, which is widely used to study the spin Hall effect, and indicates the potential of 2D TaSe$_2$ for use in spin Hall effect devices. Additionally, we have characterized the room temperature breakdown current density of TaSe$_2$ in air, where the best breakdown current density observed is $3.7\times10^7$ A/cm$^2$. This large breakdown current density further indicates the potential of TaSe$_2$ for use in 2D spin-torque devices and 2D device interconnect applications.

**Methods**

The 2D TaSe$_2$ devices were fabricated as follows. A bulk Nanosurf TaSe$_2$ sample was purchased from nanoScience Instruments and confirmed by Raman characterization shown in Figure 1(e). TaSe$_2$ flakes were prepared by the method of mechanical exfoliation using adhesive tape, depositing TaSe$_2$ flakes on an insulating substrate. For the weak anti-localization measurements, the flakes were deposited on insulating SrTiO$_3$ substrate, while, for the breakdown current measurements, flakes were deposited on 90 nm SiO$_2$ on Si substrate. Electrical contacts were defined using an electron beam lithography and liftoff process. The metal contacts were deposited by electron beam evaporation. For the weak anti-localization measurements, a 30nm/50nm Ni$_{0.8}$Fe$_{0.2}$/Cu contact was used, and for the breakdown current measurements, a 30nm/50nm Ni/Au contact was used. The weak anti-localization measurements were carried out in a $^3$He cryostat with a superconducting magnet using a Stanford Research 830 lock-in amplifier. Breakdown currents were measured at room temperature in air using a Keithley 4200 semiconductor characterization system.

The authors declare no competing financial interest.

**Acknowledgements**

This material is based upon work partly supported by NSF under Grant CMMI-1120577 and SRC under Task 2396. A portion of the low temperature measurements was performed at the National High Magnetic Field Laboratory, which is supported by National Science Foundation Cooperative Agreement No. DMR-1157490, the State of Florida, and the U.S. Department of Energy. The authors thank Z. Luo, X. Xu, E. Palm, T. Murphy, J.-H. Park, and G. Jones for experimental assistance.

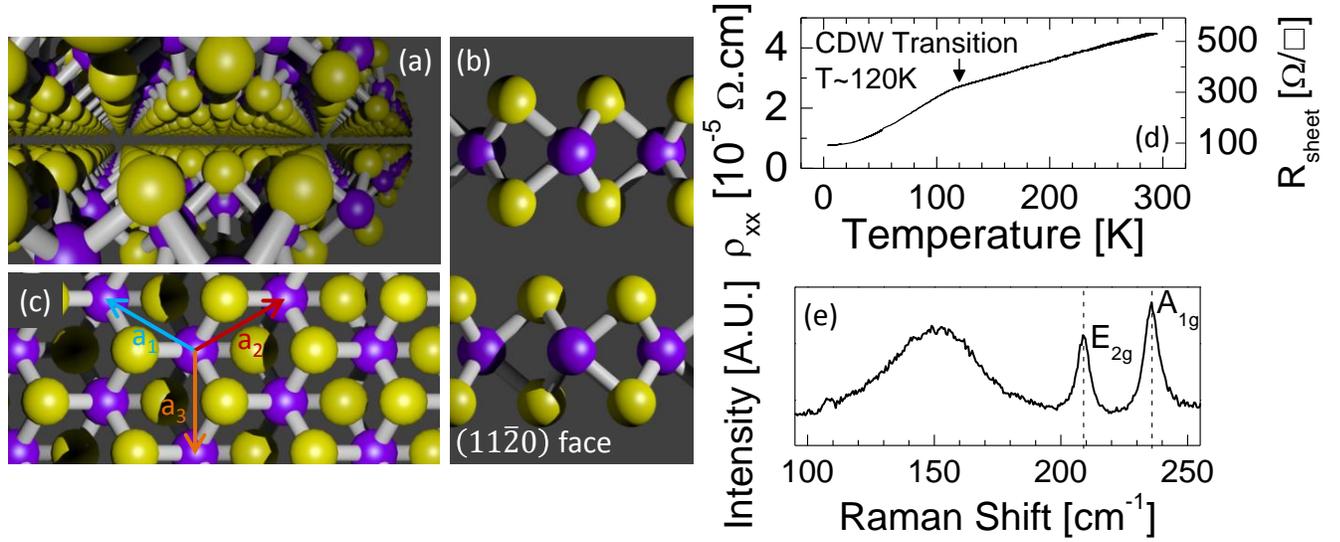

Figure 1: (a) 3D view of the 2H-TaSe$_2$ crystal structure. (b) Side view of the TaSe$_2$ crystal cleaved at the $(11\bar{2}0)$ face. (c) Top view of the TaSe$_2$ crystal with lattice vectors shown. Purple balls represent Ta atoms, while yellow balls represent Se atoms. (d) Resistivity $\rho_{xx}$ and sheet resistance $R_{sheet}$ as a function of temperate for the TaSe$_2$ flake used for weak anti-localization measurements. The change in slope at ~120K indicates that the TaSe$_2$ is of the 2H polytype. (e) Raman characterization of the bulk TaSe$_2$ from which the flakes were exfoliated.

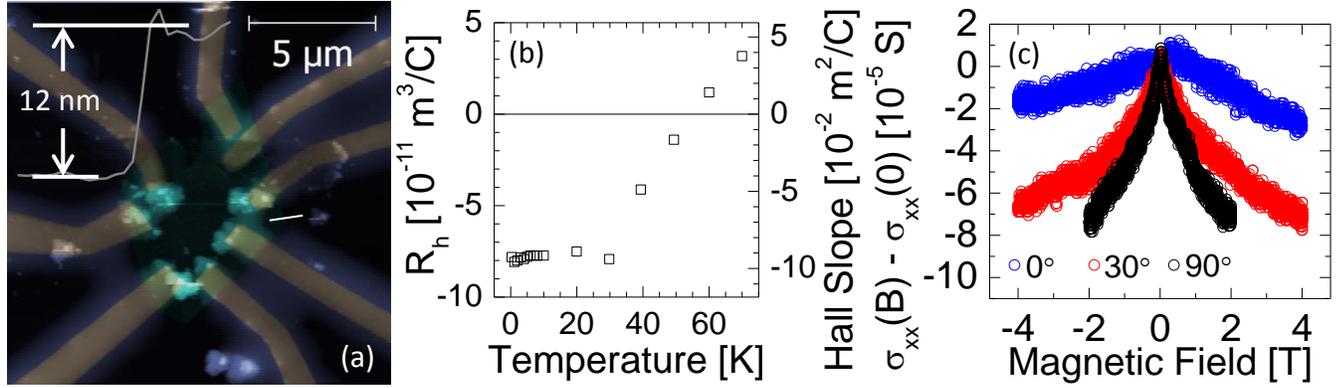

Figure 2: (a) AFM image of the TaSe$_2$ device used to study the spin-orbit scattering *via* weak anti-localization measurements. The AFM height measurement along the white line is overlaid. (b) Hall coefficient and Hall slope of TaSe$_2$ as a function of temperature. The sign change results from the reconstruction of the Fermi surface as the charge density wave state develops with decreasing temperature. (c) Differential magneto-conductivity of TaSe$_2$ at T = 0.4 K with magnetic field applied at different angles relative to the sample surface. Zero degrees indicate that the magnetic field is parallel to the TaSe$_2$ planes, while 90 degrees indicate that the magnetic field is perpendicular to the TaSe$_2$ planes.

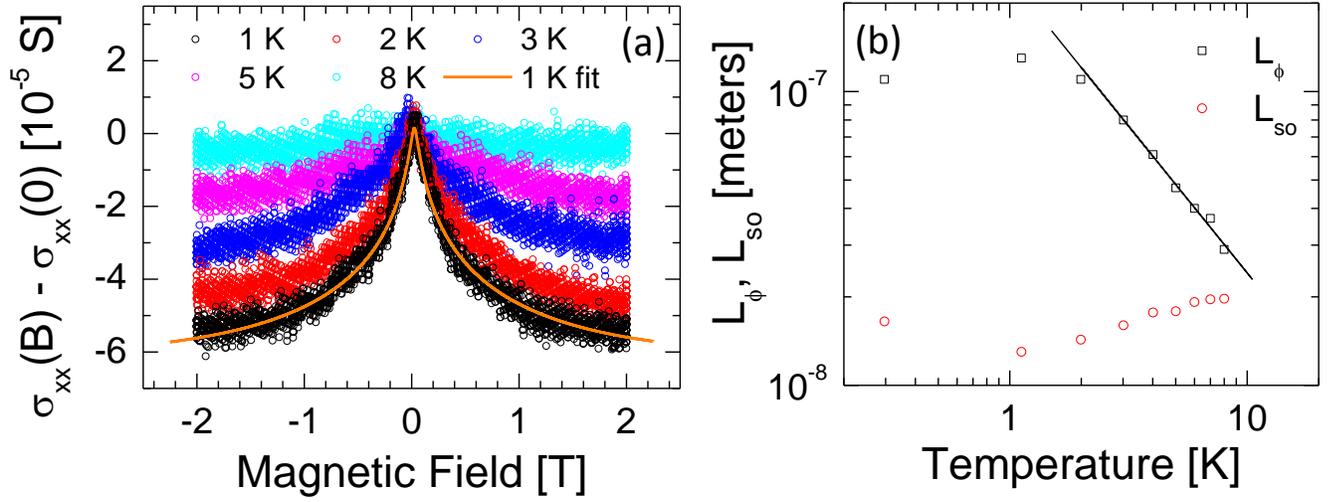

Figure 3: (a) Differential magneto-conductivity of 2H-TaSe$_2$ at temperatures from 1K to 8K. The negative magneto-conductivity shown in the figure is the characteristic of weak anti-localization and indicates the strong spin-orbit coupling of TaSe$_2$. (b) Phase coherence length L$_\phi$ (black squares) and spin-orbit scattering length L$_{so}$ (red circles) extracted from the weak anti-localization data from Figure 3(a). The black solid line indicates the T$^{-1}$ power law decrease of L$_\phi$.

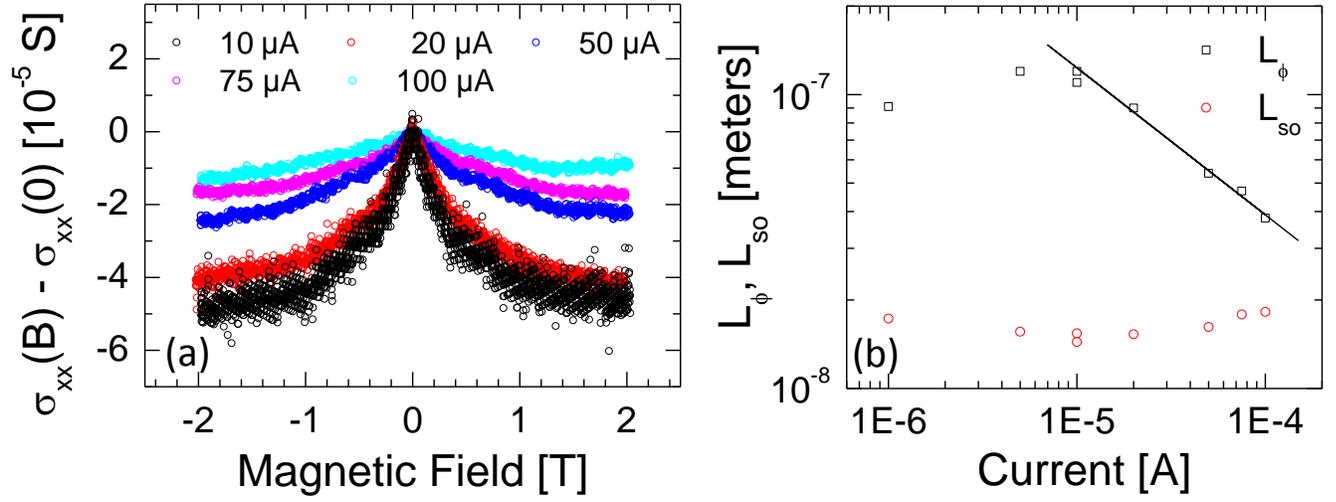

Figure 4: (a) Differential magneto-conductivity of TaSe$_2$ for different RMS bias currents as indicated in the figure. The helium bath temperature was 0.4K for these measurements. (b) Phase coherence length L$_\phi$ (black squares) and spin-orbit scattering length L$_{so}$ (red circles) extracted from the weak anti-localization data from Figure 4(a). The black solid line indicates the $I^{-1/2}$ power law decrease of L$_\phi$ with increasing bias current.

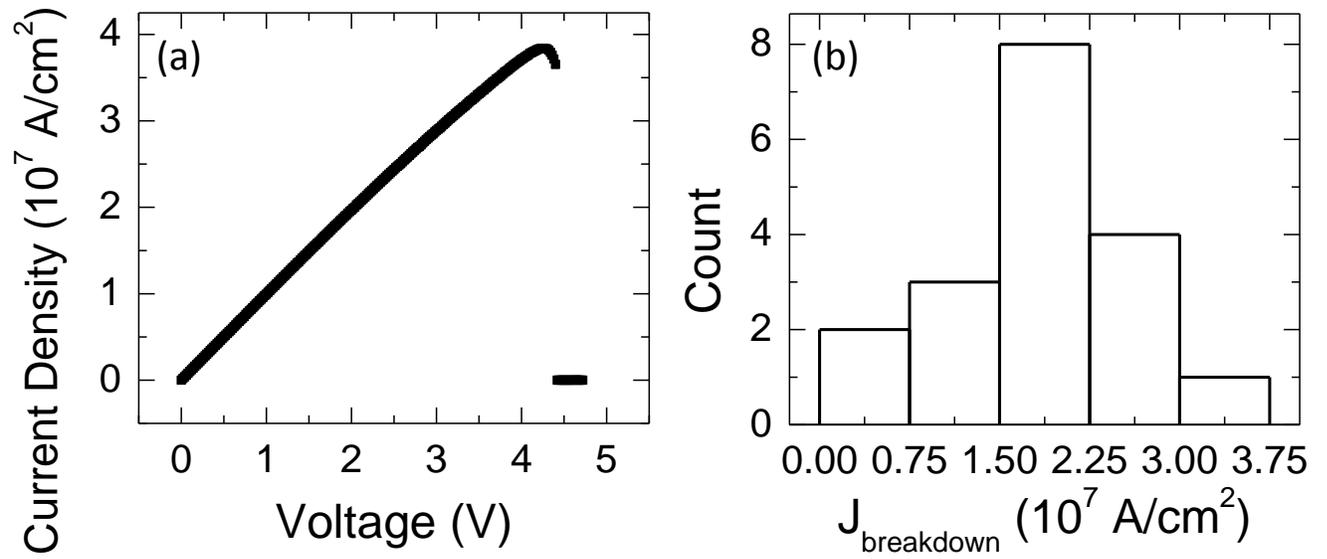

Figure 5: (a) DC current density *versus* voltage characteristic of the TaSe$_2$ device which shows the highest breakdown current. (b) Histogram of breakdown current densities for 18 TaSe$_2$ devices measured in air at room temperature.

**TOC Graphics**

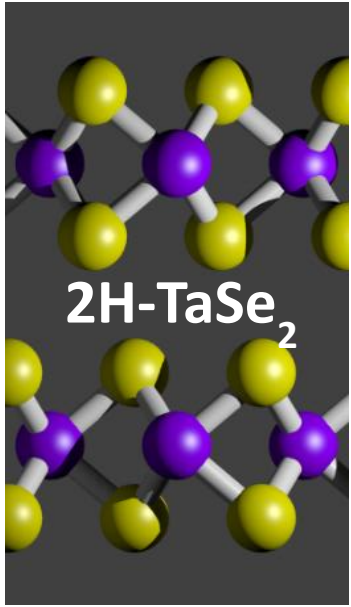
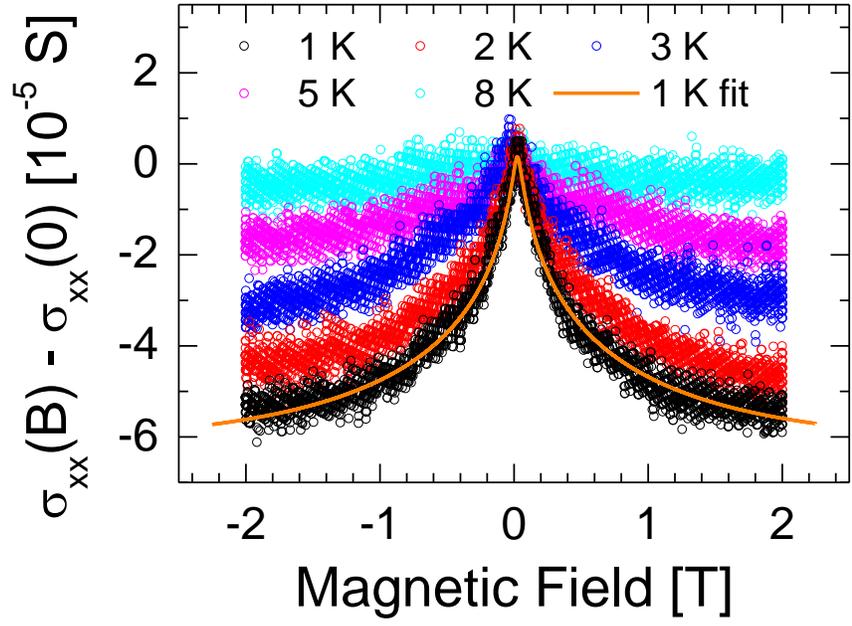